\documentclass[twocolumn,showpacs,preprintnumbers,amsmath,amssymb,floatfix]{revtex4}

\usepackage{epsf}
\usepackage{graphicx}

 \newcommand{\insertplot}[5]{\begin{figure}
 \hfill\hbox to 0.05in{\vbox to #5in{\vfill
 \inputplot{#1}{#4}{#5}}\hfill}
 \hfill\vspace{-.1in}
 \caption{#2}\label{#3}
 \end{figure}}
 \newcommand{\inputplot}[3]{
 \special{ps: plotfile #1}
}

\newcommand{\vphi}{\varphi}

\newcommand{\sqdetg}{\sqrt{-g}}

\newcommand{\rd}{{\rm{d}}}

\newcommand{\hM}{{\hat M}}
\newcommand{\hJ}{{\hat J}}
\newcommand{\vep}{{\varepsilon}}

\begin{document}

\title{
Perturbative Charged Rotating 5D Einstein-Maxwell Black Holes}
 \vspace{1.5truecm}
\author{
{\bf Francisco Navarro-L\'erida}
}
\affiliation{
{
Institut f\"ur Physik, Universit\"at Oldenburg, Postfach 2503\\
D-26111 Oldenburg, Germany}
}

\date{\today}

\begin{abstract}
We present perturbative charged rotating 5D Einstein-Maxwell
black holes with spherical horizon topology. The electric charge $Q$ is the
perturbative parameter, the perturbations being performed up to 4th
order. The expressions for the relevant physical properties of these black
holes are given. The gyromagnetic ratio $g$, in particular,
is explicitly shown to be non-constant in higher order,
and thus to deviate from its lowest order value, $g=3$.
Comparison of the perturbative analytical solutions with
their non-perturbative numerical counterparts
shows remarkable agreement.
\end{abstract}

\maketitle

{\sl Introduction}

The Kerr-Newman (KN) solution has been found to be very special in many
aspects. It represents the unique family of stationary asymptotically flat black holes
with non-degenerate event horizon of 4D Einstein-Maxwell (EM) theory. In the
general case, it represents an isolated charged rotating  black hole and it
comprises the Kerr (uncharged), Reissner-Nordstr\"om (static), and
Schwarzschild (uncharged and static) solutions as limits. 

The generalization of these black hole solutions to $D>4$
dimensions was pioneered by Tangherlini
\cite{tangher} for static black holes,
and by Myers and Perry (MP) \cite{MP} for rotating vacuum black holes.
The corresponding $D>4$ charged rotating black holes of EM theory 
could not yet be obtained in closed form \cite{MP,Horo}, although 
a subset of
solutions has been found numerically in odd dimensions \cite{numerical}.

Based on the strong interest in higher dimensional black holes in recent
years, we here take one step further towards obtaining
analytical expressions for the
higher dimensional generalizations of the KN solutions, by studying
the charged 5D MP black hole solutions perturbatively, 
solving up to 4th
order in the perturbative parameter, the electric charge.

Intriguingly, lowest order perturbation theory
gives for the gyromagnetic ratio the result $g=D-2$
\cite{Aliev}, which seems a natural higher dimensional generalization of
the gyromagnetic ratio in $D=4$ dimensions: $g=2$.
However, numerical calculations revealed that in 
higher dimensions, the gyromagnetic ratio should not be constant,
but deviate from $D-2$ for finite values of the charge.
Here we show, that in higher order perturbation theory
the gyromagnetic ratio indeed differs from $D-2$.

{\sl EM black holes}

We consider the 5D Einstein-Maxwell action with Lagragian
\begin{equation}
L = \frac{1}{16 \pi G} \sqdetg  (R - F_{\mu \nu} F^{\mu \nu}) \ , \label{lag}
\end{equation}
with curvature scalar $R$,
5-dimensional Newton constant $G$,
and field strength tensor
$
F_{\mu \nu} =
\partial_\mu A_\nu -\partial_\nu A_\mu $,
where $A_\mu $ denotes the gauge potential.

Generic stationary EM black hole solutions with 
spherical horizon topology possess two independent angular
momenta associated with two orthogonal planes of rotation \cite{MP}. 
In the case that
the two angular momenta have equal magnitude, the isometry group 
enlarges and the system of coupled Einstein and matter field equations
reduces to a system of ordinary differential equations \cite{numerical}.
For the sake of simplicity, we here focus on this case. 

To obtain perturbative charged generalizations of the 5D 
MP solutions \cite{MP}, when both angular momenta have 
equal magnitude, we employ the following parametrization
for the metric
\begin{eqnarray}
&&\rd s^2 = g _{tt} \rd t^2 + \frac{\rd r^2}{W} + r^2 (\rd \theta^2 + 
\sin^2\theta \rd \vphi_1^2 + \cos^2\theta \rd \vphi_2^2 ) \nonumber \\
&& + N ( 
\vep_1 \sin^2\theta \rd \vphi_1 + \vep_2 \cos^2\theta \rd \vphi_2 )^2 \nonumber \\
&& -2 B (\vep_1 \sin^2\theta \rd \vphi_1 + \vep_2 \cos^2\theta \rd \vphi_2 )
\rd t \label{metric} \ ,
\end{eqnarray}
and the gauge potential 
\begin{equation}
A_{\mu} \rd x^{\mu}=a_t \rd t
+a_\vphi (\vep_1 \sin^2\theta \rd \vphi_1 + \vep_2 \cos^2\theta \rd \vphi_2 )
 \ . \label{gauge_pot}
\end{equation}
The metric functions $g_{\tt}$, $W$, $N$, and $B$, 
and the functions
$a_t$, $a_\vphi$ for the gauge potential 
then depend on $r$ only. 
Here $\vep_k=\pm 1$, $k=1,2$ 
denotes the sense of rotation in the $k$-th orthogonal plane of rotation.

{\sl Perturbations}

We consider perturbations around the MP solutions, with the electric
charge as the perturbative parameter. Taking into account the symmetry with 
respect to charge reversal, the perturbations take the form
\begin{eqnarray}
&& g_{tt}=-1+\frac{2 \hM}{r^2} + q^2 g_{tt}^{(2)} + q^4 g_{tt}^{(4)} +
O(q^6) \ , \nonumber \\
&& W=1-\frac{2 \hM}{r^2}+\frac{2 \hJ^2}{\hM r^4} + q^2 W^{(2)}  + q^4
W^{(4)} + O(q^6) \ , \nonumber \\
&& N=\frac{2 \hJ^2}{\hM r^2} + q^2 N^{(2)} + q^4 N^{(4)} + O(q^6) \ ,
\nonumber \\
&& B=\frac{2 \hJ}{r^2} + q^2 B^{(2)} + q^4 B^{(4)} + O(q^6) \ ,
\nonumber \\
&& a_t=q a_t^{(1)} + q^3 a_t^{(3)} + O(q^5) \ , \nonumber \\
&& a_\vphi=q a_\vphi^{(1)} + q^3 a_\vphi^{(3)} + O(q^5) \ , 
\label{perturbations}
\end{eqnarray}
$q$ being the perturbative parameter associated with the electric charge (see
Eq.~(\ref{MJQ}) below).

When Eqs.~(\ref{perturbations}) are substituted in the system of ODE's,
obtained with the ansatz Eqs.~(\ref{metric}-\ref{gauge_pot}) 
from the field equations,
this results in a perturbative sequence of systems of
ODE's, which have to be solved order by order.

Although the systems may be solved for generic values of $\hM$ and $\hJ$, the
expressions for the metric and gauge potential perturbations are very
involved. Since the main features are shared by the extremal case we 
present here most expressions only for the extremal solutions,
while the general case will be presented elsewhere \cite{long}.

In order to perform the perturbative scheme, it is convenient to fix 
several quantities from the beginning.
In the extremal case, we have fixed the angular momentum
for any perturbative order,
and we have imposed the extremality condition for all orders. 
This choice fixes all integration constants, and it has the
advantage to allow us to compare the perturbative analytical solutions
with the non-perturbative numerical solutions 
obtained previously \cite{numerical}.

Introducing the parameter $\nu$ for the extremal MP solutions by $\hM=2\nu^2$,
$\hJ=2\nu^3$, the perturbations up to 4th order read
\begin{eqnarray*}
&&g_{tt}=-1+\frac{4\nu^2}{r^2} +\frac{r^2-4\nu^2}{3\nu^2 r^4} q^2 + \left[\frac{11 r^4 -32 \nu^2 r^2 + 16
    \nu^4}{36 \nu^6 r^6} \right.  \nonumber \\
&& \left. + \frac{4(r^2-2\nu^2)^2}{27 \nu^8 r^4} \ln
  \left(1-\frac{2\nu^2}{r^2}\right)  \right] q^4 + O(q^6) \ ,
    \nonumber \\
&&W=1-\frac{4\nu^2}{r^2}+\frac{4 \nu^4}{r^4} -\frac{r^2-2\nu^2}{3\nu^2
    r^4}q^2
    \nonumber \\
&& +\left[\frac{24 r^6 -121 \nu^2 r^4 + 181 \nu^4 r^2 -64 \nu^6}{108 \nu^8 r^6}
    \right. \nonumber \\
&&\left. +\frac{(r^2-2\nu^2)^3}{9 \nu^{10} r^4}
    \ln\left(1-\frac{2\nu^2}{r^2}\right)\right] q^4  + O(q^6) \ ,
    \nonumber \\
&&N=\frac{4\nu^4}{r^2} -\frac{2(r^2+2\nu^2)}{3 r^4} q^2  \nonumber \\
&& -\left[\frac{8 r^8 -8 \nu^2 r^6 -7\nu^4 r^4 + 16 \nu^6 r^2 -16 \nu^8}{36
    \nu^8 r^6} \right. \nonumber \\
&&+ \left. \frac{(r^2-2\nu^2 ) (3 r^6 + 8 \nu^6 )}{27 \nu^{10} r^4} \ln\left(1 
-\frac{2 \nu^2}{r^2}\right) \right] q^4  + O(q^6) \ , \nonumber \\
&&B=\frac{4 \nu^3}{r^2} - \frac{4\nu}{3 r^4}  q^2 -
    \left[\frac{(r^2-\nu^2)(r^4 -2 \nu^2 r^2 + 4 \nu^4)}{9 \nu^7 r^6}
    \right. \nonumber \\
&&+\left.  \frac{(r^2-2\nu^2)(3 r^4 -6 \nu^2 r^2 + 16 \nu^4)}{54 \nu^9
    r^4}\ln\left(1-\frac{2\nu^2}{r^2}\right) \right] q^4  \nonumber \\
&&+ O(q^6) \  ,   \nonumber \\
&& a_t=\frac{1}{r^2} q  + \left[\frac{2(r^2-\nu^2)}{9\nu^4 r^4}
    \right. \nonumber \\
&&+\left. \frac{r^2-2\nu^2}{9 \nu^6 r^2} \ln\left(1-\frac{2
    \nu^2}{r^2}\right)\right] q^3  + O(q^5) \ , \nonumber \\
\end{eqnarray*}
\begin{eqnarray}
&&a_\vphi=-\frac{\nu}{r^2}  q - \left[ \frac{2 r^4 + \nu^2 r^2 -4 \nu^4}{18
    \nu^5 r^4} \right. \nonumber \\
&&+\left. \frac{r^4-4 \nu^4}{18 \nu^7 r^2} \ln\left(1-\frac{2
    \nu^2}{r^2}\right) \right] q^3 + O(q^5) \ . \label{solution}
\end{eqnarray}

We observe that apart from the usual $1/r$ polynomial expressions, logarithms
are present. When going to higher order, more complicated structures appear
\cite{long}. 

{\sl Physical quantities}

From the analytical perturbative solutions, Eq.~(\ref{solution}), one can
extract the perturbative expressions for the physical quantities 
of these charged rotating black holes. 
Employing the same conventions as in \cite{numerical},
the mass $M$, the equal magnitude angular momenta $J$, 
and the charge $Q$ can be shown to be
\begin{eqnarray}
&&M=\frac{3}{2}\pi \nu^2 + \frac{\pi}{8\nu^2} q^2 + \frac{\pi}{288 \nu^6} 
q^4  + O(q^6) \ , \nonumber \\
&& J=\pi \nu^3 \ (\mbox{for any order}) \ , \ \ \ Q=\pi q \ , \label{MJQ} 
\end{eqnarray}
while the magnetic moment $\mu_{\rm {mag}}$ is given by
\begin{equation}
\mu_{\rm {mag}}=\pi \nu q - \frac{\pi}{18 \nu^3} q^3 + O( q^5) \
. \label{mag_mom}
\end{equation}

These perturbative extremal black holes possess an event horizon located at
$r=r_{\rm H}$, where
\begin{equation}
r_{\rm H} = \sqrt{2} \nu + \frac{\sqrt{2}}{24 \nu^3} q^2 +
\frac{11\sqrt{2}}{1152 \nu^7}  q^4 + O( q^6) \ , \label{hor_rad}
\end{equation}
which rotates with a horizon angular velocity
\begin{equation}
\Omega = \frac{1}{2\nu} - \frac{1}{24 \nu^5} q^2 - \frac{1}{288 \nu^9}
 q^4 + O(q^6) \ . \label{Omega}
\end{equation}

Introducing further the area of the horizon $A_{\rm H}$ and the electrostatic
potential at the horizon $\Phi_{\rm H}$
\begin{eqnarray}
&&A_{\rm H} = 8 \pi^2 \nu^3  + O(q^6) \ ,
\nonumber \\
&&\Phi_{\rm H}=\frac{1}{4\nu^2} q + \frac{1}{72 \nu^6} q^3 +
O(q^5) \ , \label{AH_and_PhiH}
\end{eqnarray}
one can easily see that the Smarr formula \cite{Smarr,GMT}
\begin{equation}
M=\frac{3}{2}\frac{\kappa_{\rm sg} A_{\rm H}}{8\pi G}+\frac{3}{2} 2 \Omega J +
\Phi_{\rm H} Q \ , \label{mass_formula}
\end{equation}
is satisfied up to 4th order (note,
that the surface gravity $\kappa_{\rm sg}$
vanishes for extremal solutions).

Combining Eqs.~(\ref{MJQ}-\ref{mag_mom}), 
we define the gyromagnetic ratio $g$,
\begin{equation}
g=\frac{2 M \mu_{\rm {mag}}}{Q J} = 3 + \frac{1}{12 \nu^4} q^2 +
  O(q^4) \ . \label{g_factor_extremal}
\end{equation}

In the non-extremal case (for fixed $r_{\rm H}$),
the gyromagnetic ratio $g$ obtained after lengthy
calculation is given by \cite{long}
\begin{equation}
g=3+\frac{(\hM^2 -\eta)(\hM^2+3\eta)}{\hM^2(\hM^2+\eta)^2} q^2 +
O(q^4) \ , \label{g_factor_non_extremal}
\end{equation}
where $\eta=\sqrt{\hM(\hM^3-2\hJ^2)}$. 
Note, that
Eq.~(\ref{g_factor_non_extremal}) reduces to Eq.~(\ref{g_factor_extremal}) for
extremal solutions.

{\sl Quality of the perturbative solutions}

To obtain an assessment of the quality of the perturbative solutions
we compare them with the corresponding numerical solutions \cite{numerical}.

Let us first address the black hole properties
extracted from the metric, which are obtained 
perturbatively up to 4th order in the charge.
The mass $M$ is obtained with high accuracy ($<.04\%$) up to
$Q/M \approx 0.7$, and it is still rather good ($<.3\%$) up to 
$Q/M \approx 1$,
independent of the angular momentum $J$.

This is demonstrated in Fig.~1, where we exhibit the domain
of existence of these EM black holes. 
Here the scaled angular momentum $|J|/M^{3/2}$ 
of the extremal EM black holes
is shown versus the scaled charge $Q/M$
for the exact numerical solutions and for the perturbative solutions
in 2nd and 4th order.
\begin{figure}[h!]
\begin{center}
\epsfysize=6.5cm
\mbox{\epsffile{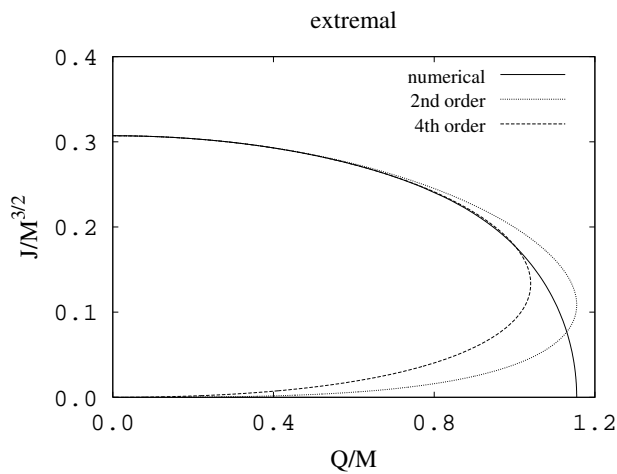}}
\caption{
Scaled angular momentum $J/M^{3/2}$ versus
scaled charge $Q/M$ for extremal black holes:
numerical (solid), 2nd order perturbation (dotted),
4th order perturbation (dashed).
}
\end{center}
\end{figure}

The horizon properties of the black holes are reproduced with
as good accuracy as the global mass.
We demonstrate this for the horizon angular velocity
in Fig.~2.
\begin{figure}[h!]
\begin{center}
\epsfysize=6.5cm
\mbox{\epsffile{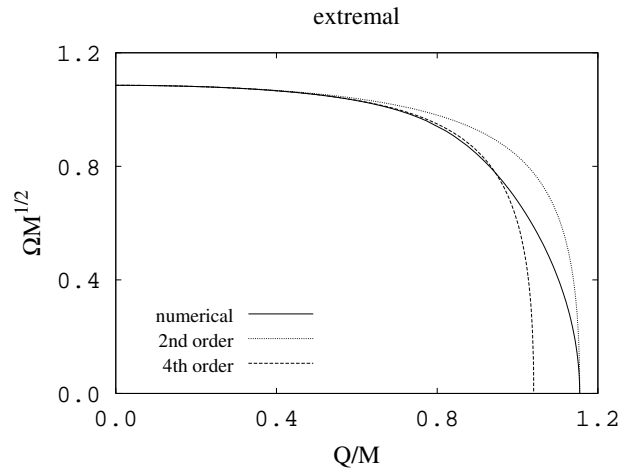}}
\caption{
Scaled horizon angular velocity $\Omega M^{1/2}$ versus
scaled charge $Q/M$ for extremal black holes:
numerical (solid), 2nd order perturbation (dotted),
4th order perturbation (dashed).
}
\end{center}
\end{figure}

The quality of the metric functions themselves is demonstrated
exemplarily in Fig.~3, 
where we show the metric coefficient $g_{tt}$
as a function of the compactified radial
coordinate $1-r_{\rm H}/r$ for an extremal black hole
with $J=5$ and $Q=3$ and compare with the perturbative. 
One can clearly see that as the order of the perturbations increases, the
agreement with the non-perturbative numerical solution improves.

\begin{figure}[h!]
\begin{center}
\epsfysize=6.5cm
\mbox{\epsffile{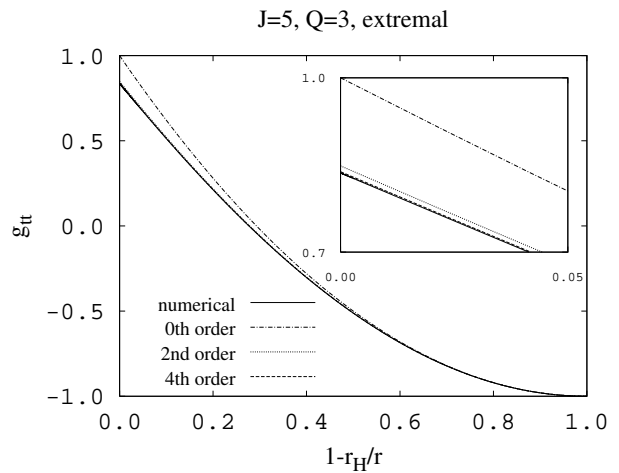}}
\caption{
Metric function $g_{tt}$ versus the compactified radial coordinate
$1-r_{\rm H}/r$ for the extremal black hole solution with angular momentum
$J=5$ and electric charge $Q=3$ ($Q/M \approx .448$):
numerical (solid), 0th order perturbation (dotted-dashed), 2nd order
perturbation (dotted), 4th order perturbation (dashed).
}
\end{center}
\end{figure}

The gyromagnetic ratio $g$, defined in Eq.~(\ref{g_factor_extremal}),
is very sensitive to the accuracy of the calculations.
Although the perturbative result in lowest order
leads to the constant value $g=3$ \cite{Aliev}, 
the higher-order perturbative calculations,
presented here, Eqs.~(\ref{g_factor_extremal}-\ref{g_factor_non_extremal}),
reveal a non-constant value for the gyromagnetic ratio.
We exhibit the gyromagnetic ratio in Fig.~4
for extremal solutions, comparing the 2nd order results
with the corresponding numerical values.
(Note, that 4th order perturbations for the gyromagnetic ratio 
require 5th order perturbations for the magnetic moment.)

\begin{figure}[h!]
\begin{center}
\epsfysize=6.5cm
\mbox{\epsffile{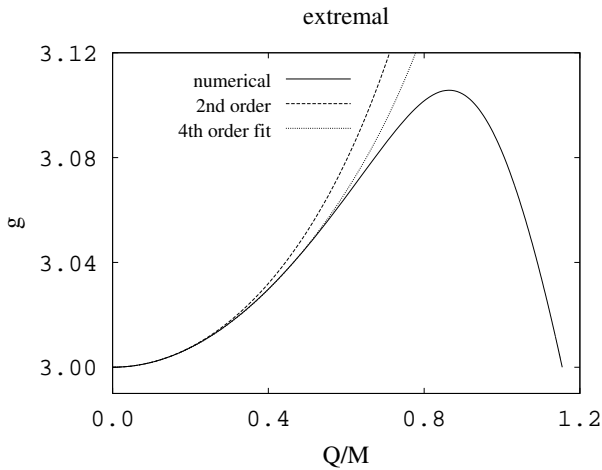}}
\caption{
Gyromagnetic ratio $g$ versus the scaled electric charge $Q/M$ for
extremal black holes:
numerical (solid), 2nd order perturbation (dashed),
4th order fit (thin-dotted).}
\end{center}
\end{figure}

As expected, for small charges the gyromagnetic ratio 
agrees very well with the numerical results.
But the accuracy holds only up to $Q/M \approx 0.2$,
since $g$ is obtained only in 2nd order, when the metric and the
gauge potential are obtained in 4th order.
Assuming a tentative 
$\nu^{-8}$ dependence for the 4th order correction
seems suited to obtain agreement 
much further, as indicated in the figure.

Similarly good or better results for $g$
are obtained for non-extremal black hole solutions. 
This is illustrated in Fig.~5, where the
gyromagnetic ratio of black holes with fixed horizon radius and 
several values of the angular momentum $J$ is shown.
\begin{figure}[t!]
\begin{center}
\epsfysize=6.5cm
\mbox{\epsffile{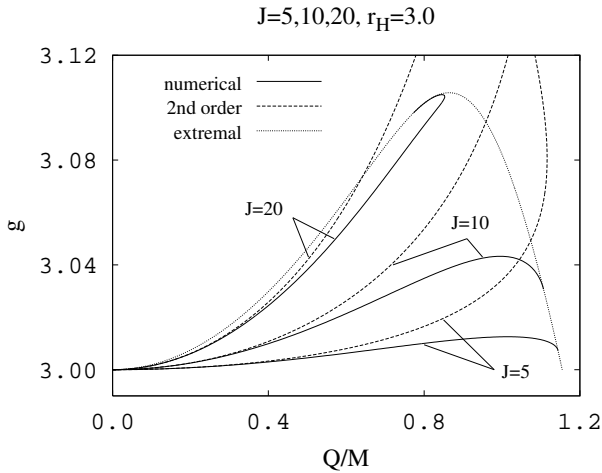}}
\caption{
Gyromagnetic ratio $g$ versus the scaled electric charge $Q/M$ for
non-extremal black holes with horizon radius $r_{\rm H}=3.0$
and angular momentum $J=5$, 10 and 20:
numerical (solid), 2nd order perturbation (dashed)
(for comparison: $g$ for extremal solutions (thin-dotted)).}
\end{center}
\end{figure}

{\sl Conclusion}

We have presented perturbative analytical solutions
for charged rotating 5D EM black holes with spherical horizon topology,
using the electric charge as the perturbative parameter. 
Contrary to the case of 4D KN black holes solutions, the
perturbative series cannot be truncated in a consistent way to produce an
exact analytical solution to the equations. 
Morever, the 4th order perturbations contain logarithms 
and in higher order more complicated structures appear \cite{long}, 
in constrast to the $1/r$ polynomial expressions of 4D KN solutions.

For the quality of the approximate perturbative solutions
we find, that the 4th order approximation is accurate
up to $Q/M \approx 0.7$ and rather good up to $Q/M \approx 1$
(recall, that $Q/M \le \sqrt{3}/2$).
Since the 4th order approximation of the metric and gauge potential functions 
gives rise to a 2nd order approximation of the gyromagnetic ratio,
this quantity is less accurate for larger values of $Q/M$.
However, the new perturbative results clearly show, that
$g \ne 3$ in general.

Although the results presented here were mainly for extremal solutions, they
are easily extended to non-extremal solutions \cite{long}.

We anticipate that this perturbative approach may be applied 
to other theories where so far only numerical solutions are available,
leading to further insight into such phenomena as
non-uniqueness, instability, counterrotation, or negative horizon mass,
as encountered for instance in Einstein-Maxwell-Chern-Simons black holes
\cite{GMT,EMCS}.

{\sl Acknowledgement}

FNL is very grateful to Jutta Kunz for valuable discussions
and comments on this paper.
FNL also gratefully acknowledges support
by the Ministerio de Educaci\'on y Ciencia 
under grant EX2005-0078 and project FIS2006-12783-C03-02.

\end{document}